\documentclass[preprint]{aastex631}

\usepackage{ulem}
\usepackage{multirow}
\usepackage{longtable}
\usepackage{graphicx}

\newcommand{\feh} {\mbox{\rm [Fe/H]}}

\newcommand{\mgfe} {\mbox{\rm [Mg/Fe]}}

\newcommand{\U} {\mbox{\rm $U$}}
\newcommand{\W} {\mbox{\rm $W$}}
\newcommand{\VR} {\mbox{\rm $V_{\rm{R}}$}}
\newcommand{\Vz} {\mbox{\rm $V_{\rm{Z}}$}}
\newcommand{\Vphi} {\mbox{\rm $V_{\rm{\phi}}$}}
\newcommand{\kmprs} {\mbox{ ${\rm km}\, {\rm s}^{-1}$\,}} 
\newcommand{\kmprskpc} {\mbox{ ${\rm km}\, {\rm s}^{-1}$\, kpc}} 
\newcommand{\kpcprdeg} {\mbox{ ${\rm kpc}\, {\rm deg}^{-1}$\,}}

\newcommand{\Rapo} {\mbox{\rm $R_{\rm apo}$}}
\newcommand{\Rper} {\mbox{\rm $R_{\rm per}$}}

\newcommand{\Rg} {\mbox{\rm $R_{\rm g}$}}
\newcommand{\Lz} {\mbox{\rm $L_{\rm z}$}}

\begin{document}

\title{Evidence for Corotation Origin of Super Metal-Rich Stars in LAMOST-Gaia: 
Multiple Ridges with a Similar Slope in $\phi$ versus $\Lz$ Plane}


\author[0000-0002-8442-901X]{Yuqin Chen}
\altaffiliation{CAS Key Laboratory of Optical Astronomy, National Astronomical Observatories, Chinese Academy of Sciences, Beijing 100101, China;cyq@nao.cas.cn,gzhao@nao.cas.cn}
\altaffiliation{School of Astronomy and Space Science, University of Chinese Academy of Sciences, Beijing 100049, China}
\author[0000-0002-8980-945X]{Gang Zhao}
\altaffiliation{CAS Key Laboratory of Optical Astronomy, National Astronomical Observatories, Chinese Academy of Sciences, Beijing 100101, China;cyq@nao.cas.cn,gzhao@nao.cas.cn}
\altaffiliation{School of Astronomy and Space Science, University of Chinese Academy of Sciences, Beijing 100049, China}
\author[0000-0003-3265-9160]{Haopeng Zhang}
\altaffiliation{CAS Key Laboratory of Optical Astronomy, National Astronomical Observatories, Chinese Academy of Sciences, Beijing 100101, China;cyq@nao.cas.cn,gzhao@nao.cas.cn}
\altaffiliation{School of Astronomy and Space Science, University of Chinese Academy of Sciences, Beijing 100049, China}

\begin{abstract}
Super metal-rich (SMR) stars in the solar neighborhood are thought to be born in
the inner disk and came to present location by radial migration, which is most intense at the co-rotation resonance (CR) of the Galactic bar. In this work, we show evidence for the CR origin of SMR stars in LAMOST-Gaia by detecting six ridges and undulations in the $\phi$ versus $\Lz$ space coded by median $\VR$, following a similar slope of $-8 \kmprs \kpcprdeg$. The slope is predicted by Monario et al.'s model for CR of a large and slow Galactic bar.
For the first time, we show the variation of angular momentum with  azimuths from $-10^{\circ}$ to $20^{\circ}$ for two outer and broad undulations with negative $\VR$ around $-18 \kmprs$ following this slope. The wave-like pattern with large amplitude outside CR and a wide peak of the second undulations indicate that minor merger of the Sagittarius dwarf galaxy with the disk might play a role besides the significant impact of CR of the Galactic bar.
\end{abstract}



\section{INTRODUCTION}
With the release of the Gaia data, rich structures in the phase-space distribution have been revealed. For instance, multiple ridges displayed in velocity space were found by \cite{2018A&A...616A..11G} and \cite{2018A&A...619A..72R}. \cite{2019MNRAS.488.3324F} also reported the ridge of the Hercules moving group, and many features, so-called "horn" and "hat", in the $R$ versus $\Vphi$ coded by mean $\VR$ velocity. The origins of these rich structures is a debate topic with the Hercules moving group as a typical case.
It has been suggested that the Hercules moving group, {first reported in \cite{1999ApJ...524L..35D}, was
formed outside of the bar’s outer Lindblad resonance (OLR) based on a faster bar \citep{2016MNRAS.457.2569M}.
However, \cite{2017ApJ...840L...2P} favored for the scenario that
orbits trapped at the co-rotation resonance (CR) of a slow bar could produce the Hercules moving group in local velocity space.  Based on the Galactic model of \citet{2017ApJ...840L...2P}, \citet{2019A&A...626A..41M} proposed that no fewer than six ridges in local action space that can be related to resonances of this slow bar, which induces a wave-like pattern with wavenumber of $m=6$ excitation. However, \cite{2019MNRAS.490.5414F} explained the ridges in the average Galactocentric radial velocity as a function
of angular momentum and azimuth as a wavenumber $m=4$ pattern caused by spiral arms.

Most of the observed structures in the phase-space distribution can be explained by different combinations of non-axisymmetric perturbations, making their modeling degenerate \citep{2019MNRAS.490.1026H}. Meanwhile, the relative contribution of the CR and OLR resonances (e.g. for the Hercules moving group) is different between the slow and fast rotation speed of the Galactic bar.  The combination of different resonances due to various perturbations make it difficult to discover their origins.
Following the ridges as a function of azimuth provides a promising way to disentangle the effect of different resonances. In this respect, \citet{2019A&A...632A.107M} shows that the Hercules angular momentum changes significantly with azimuth as expected for the CR resonance of a dynamically old large bar. They proposed that such a variation would not happen close to the OLR of a faster bar at least for 2 Gyr after its formation.

In this letter, we investigate the variation of angular momentum ($\Lz$) with Galactic azimuth ($\phi$) coded by $\VR$ using SMR stars in the LAMOST-Gaia survey as tracers. Since SMR stars in the solar neighborhood are thought to originate from the inner disk \citep{2015A&A...582A.122K,2019AJ....158..249C,2003ApJ...591..925C}, features related to the CR resonance of the Galactic bar should be easily identified in this special population.

\section{Data}
SMR stars with $\feh>0.2$ and spectral
signal-to-noise ratio (S/N) larger than 10 are selected from LAMOST DR7 \citep{2006ChJAA...6..265Z,2012RAA....12..723Z,2012RAA....12.1197C,2012RAA....12..735D,2015RAA....15.1089L}, which provide radial velocity and updated stellar parameters based on methods in \cite{2015RAA....15.1095L}.
Then we cross-match this sample  with {\it Gaia} EDR3 \citep{2021A&A...649A...1G} to obtain proper motions, 
and limit stars with errors in proper motions less than 0.05 $mas/yr$, and
the renormalized unit weight error of $RUWE<1.44$ \citep{2018A&A...616A...2L}.
Bayesian distances are from the StarHorse code \citep{2019A&A...628A..94A} using {\it Gaia} EDR3 and stars with relative error less than 10\% are adopted. The  radial velocity are based on the LAMOST survey, and stars with errors in radial velocity larger than 10 $\kmprs$ are excluded from the sample.

Galactocentric positions, spatial velocity and orbital parameters (apocentric/pericentric distances, $\Rapo$ and $\Rper$) are calculated 
based on the publicly available code {\it Galpot} 
with the default potential ({\it MilkyWayPotential}) provided by 
\cite{2017MNRAS.465...76M}. Note that $\Lz$ are calculated with the definition of Equation (3.72) as presented in \cite{2008gady.book.....B}, while \citet{2019A&A...632A.107M} adopted an approximate value of $R \Vphi$.  
We use the cylindrical coordinate ($\VR$, $\Vphi$, $\Vz$) with the Sun's distance
from the Galactic centre  $R=8.21$ kpc, the solar peculiar velocity of ($U_{\odot}, V_{\odot}, W_{\odot}) 
= $ (11.1, 12.24, 7.25) $\kmprs$ \citep{2010MNRAS.403.1829S} and the circular speed of $V_{c} = 233.1 \kmprs$ \citep{2011MNRAS.414.2446M}.
The sample have 214,199 star stars with velocities between $-300 \kmprs$ and $+300 \kmprs$, and the velocity dispersions of (36.1,23.1, 17.3)  $\kmprs$ in ($\VR$, $\Vphi$, $\Vz$). Their Galactic locations are of $7<R<10$ kpc, and $-0.5<Z<1.5$ kpc with peaks at $R=8.5$ kpc and $|Z|\sim 0.25$ kpc, respectively.

\section{The $\phi$ versus $\Lz$ plane}
Figure 1 shows the $\phi$ versus $\Lz$ space coded by median $\VR$ for SMR stars. There are six ridges and undulations, generally following a similar slope of  $-8 \kmprs \kpcprdeg$, which is predicted for stars with orbits trapped at CR in the Galactic model of \citet{2019A&A...626A..41M}.  The red ridge with positive median $\VR$ following the red line of $Lz=-8*\phi+1600 \kmprskpc$ corresponds to the Hercules moving group in \citet{2019A&A...632A.107M}, and for comparison we plot the green line, which has a zero slope with $\Lz=1600 \kmprskpc$ at $\phi=0 $, as expected for the OLR in the bar model. Note that the range of mean $\VR$ in \citet{2019A&A...632A.107M} (their Fig.~2) is of $\pm 20 \kmprs$, while the {\bf coverage} of $\pm 50 \kmprs$ in median $\VR$ is {adopted} in our Fig.~1. The uncertainties of the median $\VR$ for these features are of 0.5-1.0 km/s.

Except for the Hercules moving group, two 
ridges (with intercepts of $1380$ and $1920$ $\kmprskpc$) and three undulations (with intercepts of $1500$, $1800$ and $2100$ $\kmprskpc$) are clearly shown.
The similar slope in the $\Lz$ versus $\phi$ plane for the six
ridges and undulations indicates that the bar's resonance is not limited within the location of Hercules moving group ($\Lz=1600 \kmprskpc$), but has significant effect in the solar neighborhood ( $\Lz=2100 \kmprskpc$).

Meanwhile, positive median $\VR$ are found for the inner region of $\Lz<1700 \kmprskpc$, above which median $\VR$ are negative except for a narrow ridge of $Lz=-8*\phi+1920 \kmprskpc$ with median $\VR$ in the order of $10 \kmprs$. This transition may indicates that minor merge event might start to take effect, which significantly increase the variation amplitude and make the undulations wider. Since the slope of  $-8 \kmprs \kpcprdeg$ persists, we expect that the role of the CR is still significant in the outer region of $\Lz>1700 \kmprskpc$ for SMR stars. Interestingly, \cite{2022arXiv220606207G} 
found a somewhat similar
velocity distribution with a lower velocity variation (see their Fig. 16), but a change in the sign of $\VR$ velocity (as a function of Galactocentric angle) at a radius of 10 kpc, in phase with the bar angle, is not seen in our SMR sample because only a small fraction of SMR stars in our sample could reach to 10 kpc.

Although the wave-like pattern between the ridges and undulations are found for both the inner and outer regions but the amplitudes and the widths are different. The two blue undulations in the outer region have (negative) median $\VR$ of $\sim -18 \kmprs$, while the two ridges in the inner disk have (positive) median $\VR$ around $21 \kmprs$. The width of the outest undulation at $\Lz=2100 \kmprskpc$ is largest, spanning a range of 2.5 kpc from $7$ to $9.5$ kpc in term of guiding radius $\Rg$, and the median $\VR$ is the most negative, which indicates extra mechanisms, such as the minor merge of the Sgr dwarf galaxy, may play an important contribution to this undulation. 

\begin{figure*}
\includegraphics[scale=0.8]{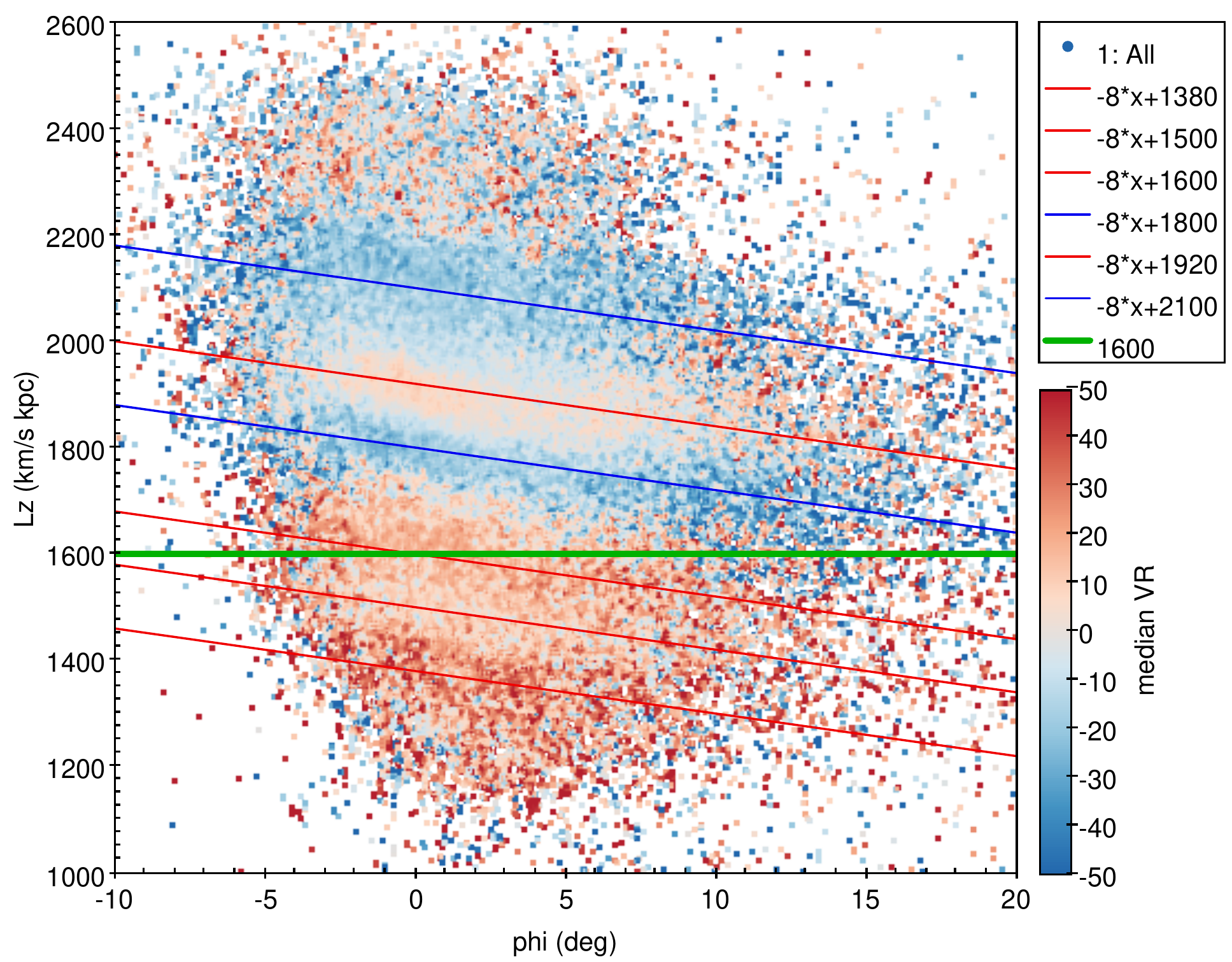} 

\caption{Median $\VR$ in the $\phi$ versus $\Lz$ space for SMR stars.The red lines for ridges (high $\VR$) and blue lines for undulations (low $\VR$) correspond to a slope of $-8 \kmprs \kpcprdeg$ (expected for CR), while the green line corresponds to
 $0 \kmprs \kpcprdeg$ (expected for OLR). Note that the red line of $\Lz=-8 \phi+1600 \kmprskpc$ corresponds to the location of the Hercules moving group as previously reported by \citet{2019A&A...632A.107M}.}
\label{f1}
\end{figure*}

\section{The distributions of pericentre and apocentre distances}
In order to know how far these SMR stars can reach in the Galactic inner and outer disks,
Fig.~2 shows the distribution of pericentre and apocentrc distances for all stars (red solid lines) and stars aligned with the Sun and the Galactic centre (black dash lines).
Then we fit the distributions with two gaussian functions for stars at the solar azimuth ($\phi=0$) and obtain two peaks for pericentre and apocentre distances respectively.
The main peak of pericentre distance is at 7.25 kpc and a second peak at 5.62 kpc, while the histogram of apocentre distance shows a main peak at 8.6 kpc and the second peak at 9.4 kpc. It is found that 79\% SMR stars excurse within 4 kpc around the location of the Sun at 8.2 kpc. The second peak of pericentre distance at 5.6 kpc is interesting because it is close to the CR position at 6 kpc in  \citet{2017ApJ...840L...2P} and very close to 5.5 kpc in \cite{2019MNRAS.490.4740B} for a slow bar.
There are 36\% SMR stars with pericentre distance less than 6.5 kpc, and they are significantly affected by the CR of the Galactic bar. Finally, the apocentre distribution has a second peak at 9.4 kpc, which is within the the bar's OLR at about 10.5 kpc according to \citet{2017ApJ...840L...2P}.

\begin{figure*}
\includegraphics[scale=0.8]{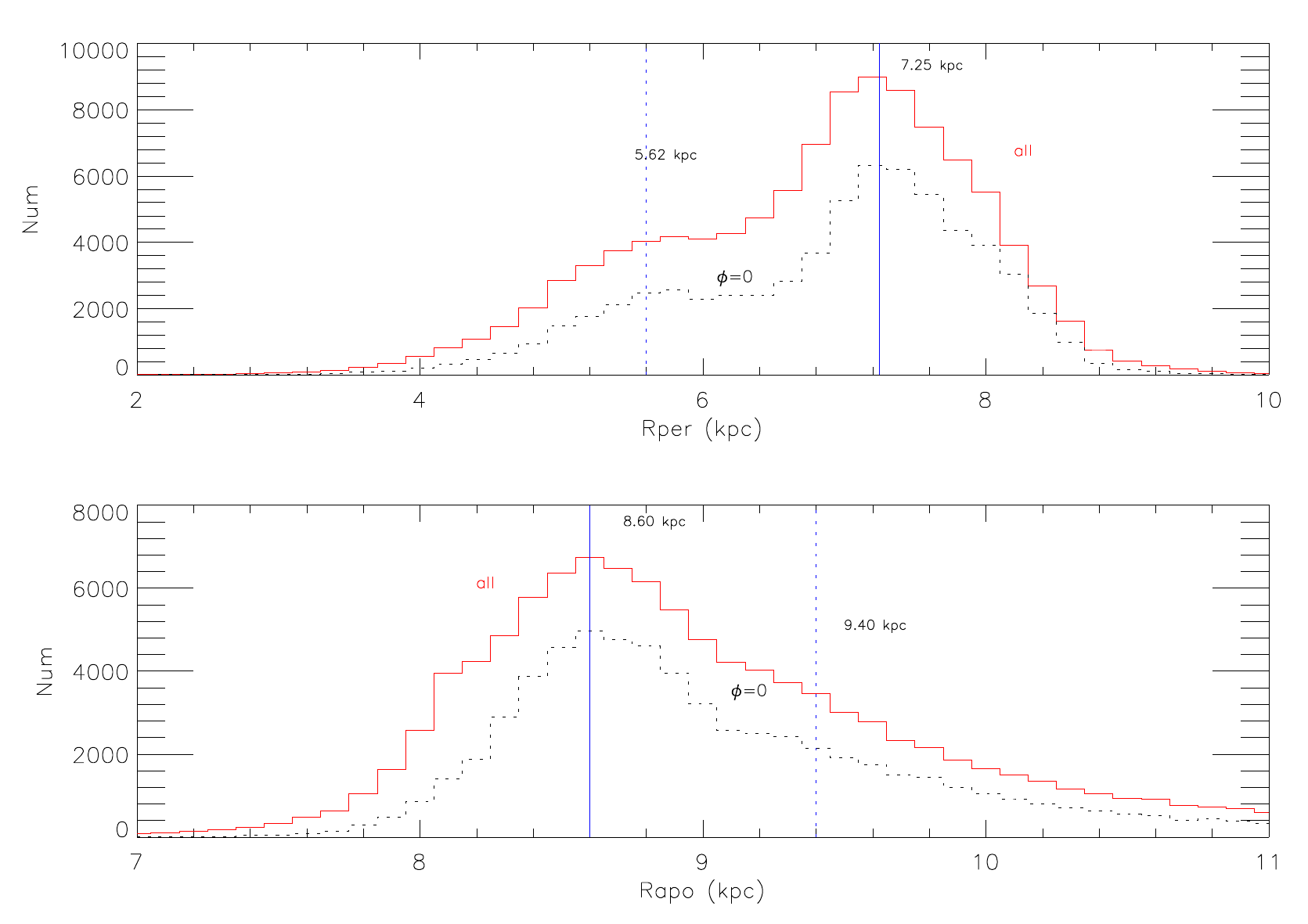}
\caption{The distribution of pericentre and apocentre distances for all SMR stars (solid lines) and stars at the solar azimuth of $\phi\sim0$ (dash lines).
	The blue lines are for the main (solid) and the second (dot) peaks in the distribtuions.}
\label{f2}
\end{figure*}

\section{Comparison with the Galactic bar model}

Based on a realistic model for a slowly rotating large Galactic bar with pattern speed
of $\Omega_b=39 \kmprs\, kpc^{-1}$, \citet{2019A&A...626A..41M} show no fewer than six ridges in local action space that can be related to resonances with the bar. It is interesting that SMR stars in the present sample do show six ridges and undulations in the $\Lz$ versus $\phi$ plane. For direct compariosn, we adopt the Galactic coordinate frame for stellar velocity (U,V,W) and show the $R-V$ diagram coded by median $-U$ in Fig.~3, which matches their Fig.~6 quite well. Specifically, the ridge at  $\Lz$ of $1380$ corresponds to CR (their green line) feature and the undulation at  $\Lz=2100 \kmprskpc$ fits the OLR (their red line) feature. The two ridges (at $1600$ and $1920 \kmprskpc$) are associated with the 6:1 (pink) and 3:1 (purple) resonances, and one undulation at $1800 \kmprskpc$ is related to the 4:1 (blue) resonance.

Note that the two ridges at  $\Lz=1380 \kmprskpc$ and  $\Lz=1600 \kmprskpc$ in the present work are the same as
the strong positive $\VR$ features near  $\Lz=1400$ and  $\Lz=1600$ observed 
in the solar neighborhood by \citet{2021MNRAS.505.2412C} based on {\it Gaia DR2} as a result of the bar's resonance. They also suggested a slow bar with the current pattern speed of  $\Omega_b=35.5 \kmprs\, kpc^{-1}$, and placed the corotation
radius at $6.6$ kpc. Moreover, \citet{2021MNRAS.500.4710C} introduced a decelerating bar model, which can reproduce with
its corotation resonance the offset and strength of the Hercules stream in the local $\VR$ versus $\Vphi$ plane and the double-peaked structure of mean $\VR$ in the $\Lz - \phi$ plane due to the accumulation of orbits near the boundary of the resonance. Further work on the comparison of the model's result with observation is desire.

\begin{figure*}
\includegraphics[scale=0.8]{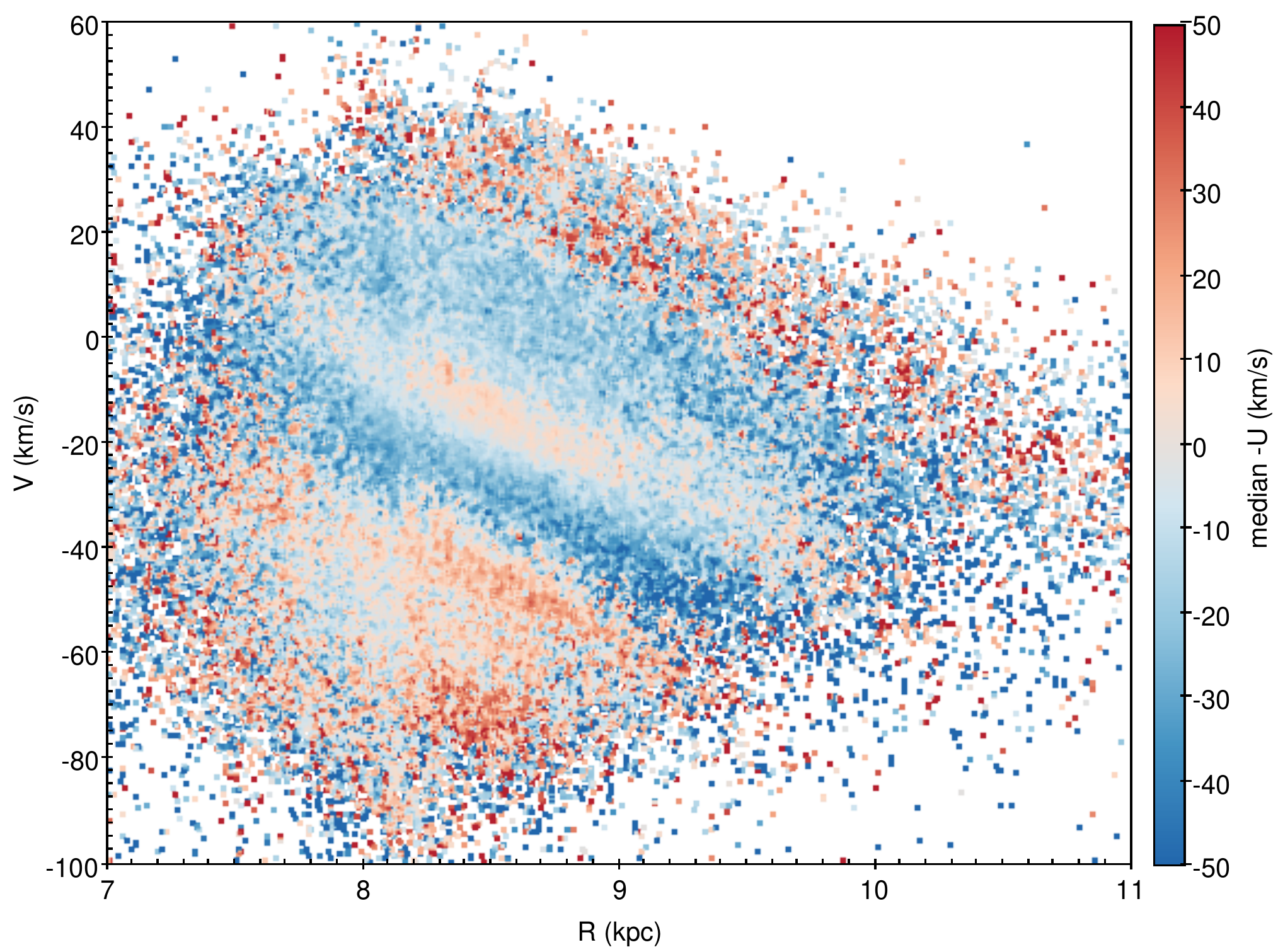}
\caption{The $R$ versus $V$ coded by median $-U$ for SMR stars. There is also a similar slope found for the six ridges and undulations.}
\label{f3}
\end{figure*}

In sum, the multiple ridges and undulations found for SMR stars in Fig.~1 can be explained by the bar resonances in the model of \citet{2019A&A...626A..41M}. Their similar slope indicates that these features (even in the OLR region) are affected by the CR of the slow and long bar. 
But the strong $\VR$ modulations from ridges to undulations and very wide range in the last undulation beyond the CR region suggest
that minor merger} may also play an role in the Galactic disk.  

Finally, we investigate chemical signatures of the six ridges and undulations based on abundances in \cite{10.1093/mnras/stac1959}, and there are 42,109 SMR stars with $\mgfe$ ratios available. There is no difference in $\mgfe$ among ridges and undulations. Both have a peak at $\mgfe \sim 0.05$ dex, typical for the old bar. Based on the LAMOST middle resolution survey, 
\cite{2021RAA....21..153Z} also suggested that SMR stars have slightly enhanced $\mgfe=0.08$ dex. Note that stars from the Sgr dwarf galaxy itself usually have low $\mgfe\sim-0.05$ at $\feh\sim-0.4$, and there is no star with solar metallicity as shown in Fig.~9 of \cite{2021SCPMA..6439562Z}. Therefore, these SMR stars in the present work are not from the Sgr dwarf galaxy itself, but the minor merge event of the Sgr dwarf galaxy with the Galactic disk could induce strong modulations of $\VR$ from ridges to undulations, and make the undulation at $\Lz=2100 \kmprskpc$ becomes wider.

\section{Conclusions}
We have detected, for the first time, six ridges and undulations following a single slope
in the $\phi$ versus $\Lz$ plane coded by median $\VR$ from a specific population of SMR stars based on LAMOST DR7 and {\it Gaia} EDR3.
Specifically, the variation of radial velocity with angular momentum $\Lz$ and azimuth $\phi$ for the six ridges and undulations is seen with a similar slope of $-8 \kmprs \kpcprdeg$, which is predicted for stars with orbits trapped at the CR of a slow bar in the model of \citet{2019A&A...626A..41M}.
The median $\VR$ shifts from positive to negative values at $\Rg\sim 7.4$ kpc (for
$\phi=0$, $\Lz\sim1700 \kmprskpc$).
This transition may indicate the role of minor merge starts to take effect together with the contribution from the CR of the bar. The most outer undulation around $\Rg\sim 8.7$ kpc (for
$\phi=0$, $\Lz\sim2100 \kmprskpc$) has a wide feature (three times larger than ridges), which is probably also related to the minor merge event of the Galactic disk with Sgr dwarf galaxy, but this
remains an open question for further study in the future. Moreover, since the major merge event by the Gaia-Sausage-Enceladus galaxy bring its metal-rich component \citep{2021SCPMA..6439562Z}, and the accreted halo stars with special chemistry \citep{2019NatAs...3..631X}, into the solar neighborhood, it is interesting to probe how major merge events take effect
on the existence of these ridges and undulations. Finally, many moving groups exist in the solar neighborhood \citep{2009ApJ...692L.113Z}, and it is of high interest to probe if they may leave some imprints in the $\phi$ versus $\Lz$ plane coded by median $\VR$ as the Hercules moving group.

\begin{acknowledgements}
This study is supported by the National Natural Science Foundation of China under Grant Nos. 11988101, 11890694, and National Key R\&D Program of China under Grant No. 2019YFA0405502.\\
Guoshoujing Telescope (the Large Sky Area Multi-Object Fiber Spectroscopic Telescope, LAMOST) is a National Major Scientific Project has been provided by the National Development and Reform Commission. LAMOST is operated and managed by the National Astronomical Observatories, Chinese Academy of Sciences. This work has made use of data from the European Space Agency (ESA) mission {\it Gaia} (\url{https://www.cosmos.esa.int/gaia}), processed by the {\it Gaia} Data Processing and Analysis Consortium (DPAC, \url{https://www.cosmos.esa.int/web/gaia/dpac/consortium}). Funding for the DPAC has been provided by national institutions, in particular the institutions participating in the {\it Gaia} Multilateral Agreement.
\end{acknowledgements}

\bibliography{msv1}{}

\bibliographystyle{aasjournal}

\end{document}